\documentclass{PoS}

\usepackage{epsfig}

\PoS{PoS(LAT2005)335}

\title{Non-degenerate quark mass effect on $B_K$ with a mixed action }

\ShortTitle{Non-degenerate quark mass effect on $B_K$ with a mixed action}

\author{\speaker{Taegil Bae} \\ Department of Physics, Seoul National
University, Seoul, 151-747, South Korea \\ E-mail:
\email{esrevinu@phya.snu.ac.kr}}

\author{Jongjeong Kim \\ Department of Physics, Seoul National
University, Seoul, 151-747, South Korea \\ E-mail:
\email{rvanguard@phya.snu.ac.kr}}

\author{Weonjong Lee \thanks{This research is supported in part by the
KOSEF grant (R01-2003-000-10229-0), by the MOST/KISTEP grant of
international collaboration, and by the BK21 program.} \\ Center for
Theoretical Physics, Department of Physics, Seoul National University,
Seoul, 151-747, South Korea \\ E-mail: \email{wlee@phya.snu.ac.kr}}

\abstract{ We investigate the effect of non-degenerate quarks on
$B_K$. This effect is noticeably large for $B_K$ (significantly larger
than statistical uncertainty). Hence, it is important to include this
effect in order to determine $B_K$ with higher precision. We also
observe that the quality of fitting for $B_K$ gets better when we
include non-degenerate combinations to fit to the prediction by Van de
Water and Sharpe. However, this effect on $B_7^{(3/2)}$ and
$B_8^{(3/2)}$ turns out to be relatively small.  }

\FullConference{XXIIIrd International Symposium on Lattice Field Theory\\
25-30 July 2005\\
Trinity College, Dublin, Ireland}

\begin{document}
\section{Effect of non-degenerate quarks on $B_K$}
\label{sec:bk}
This paper is a follow-up from the previous paper \cite{ref:wlee:10}
on $B_K$. 
Hence, in order to avoid repeating the same explanation, we will quite
often refer to Ref.~\cite{ref:wlee:10} for details.
%

In this paper, we focus on $B_K$, the kaon bag parameter for indirect
CP violation. In particular, we are interested in the effect of
non-degenerate quarks ({\em i.e.} $m_x \ne m_y$).
Before we go into the details, we want to address couple of issues
which covers the historical evolution of the staggered
$\epsilon'/\epsilon$ project.
When we used unimproved staggered fermions to calculate weak matrix
elements, we have observed three major problems: (1) large scaling
violation, (2) large perturbative corrections at the one loop level,
and (3) large uncertainty due to the quenched approximation.
We use improved staggered fermions with HYP fat links which reduce
the scaling violations and decrease the one-loop correction down
to $\approx$ 10\% level \cite{ref:wlee:1,ref:wlee:2,ref:wlee:3}.
We measure weak matrix elements over a subset of the MILC gauge
configurations in order to remove the uncertainty originating from
quenched approximation \cite{ref:milc:1}.
In other words, we numerically simulate the unquenched QCD where the
valence quarks are HYP staggered fermions and the sea quarks are
AsqTad staggered fermions (we call this a ``mixed action'').
Hence, in this paper, we report recent progress in calculating $B_K$
in unquenched QCD using improved staggered fermions.
%

Details of input parameters for this numerical study are provided in
Ref.~\cite{ref:wlee:10} and so we do not repeat them here.
We use 5 degenerate quark combinations and 4 non-degenerate
combinations to probe the effect of non-degenerate quarks.
These 9 combinations are listed in Table \ref{tab:bk}.
Note that there are two pairs of which the degenerate and
non-degenerate combinations share the same kaon mass at the leading
order (for example, $m_x = m_y = 0.03$ and $m_x=0.01$ \& $m_y=0.05$).
\begin{table}[h!]
\begin{center}
\begin{tabular}{| c | c || c |}
\hline
$m_x$ & $m_y$ & $B_K$ \\
\hline \hline
0.01 & 0.01 &  0.6108 $\pm$ 0.0164 \\
0.02 & 0.02 &  0.6421 $\pm$ 0.0065 \\
0.03 & 0.03 &  0.6722 $\pm$ 0.0039 \\
0.04 & 0.04 &  0.6955 $\pm$ 0.0029 \\
0.05 & 0.05 &  0.7144 $\pm$ 0.0024 \\
\hline
0.01 & 0.05 &  0.6869 $\pm$ 0.0068 \\
0.02 & 0.05 &  0.6907 $\pm$ 0.0039 \\
0.03 & 0.05 &  0.6976 $\pm$ 0.0030 \\
0.04 & 0.05 &  0.7058 $\pm$ 0.0027 \\
\hline
\end{tabular}
\end{center}
\caption{Quark mass combinations and corresponding $B_K$ values}
\label{tab:bk}
\end{table}

\begin{figure}[h!]
\epsfig{file=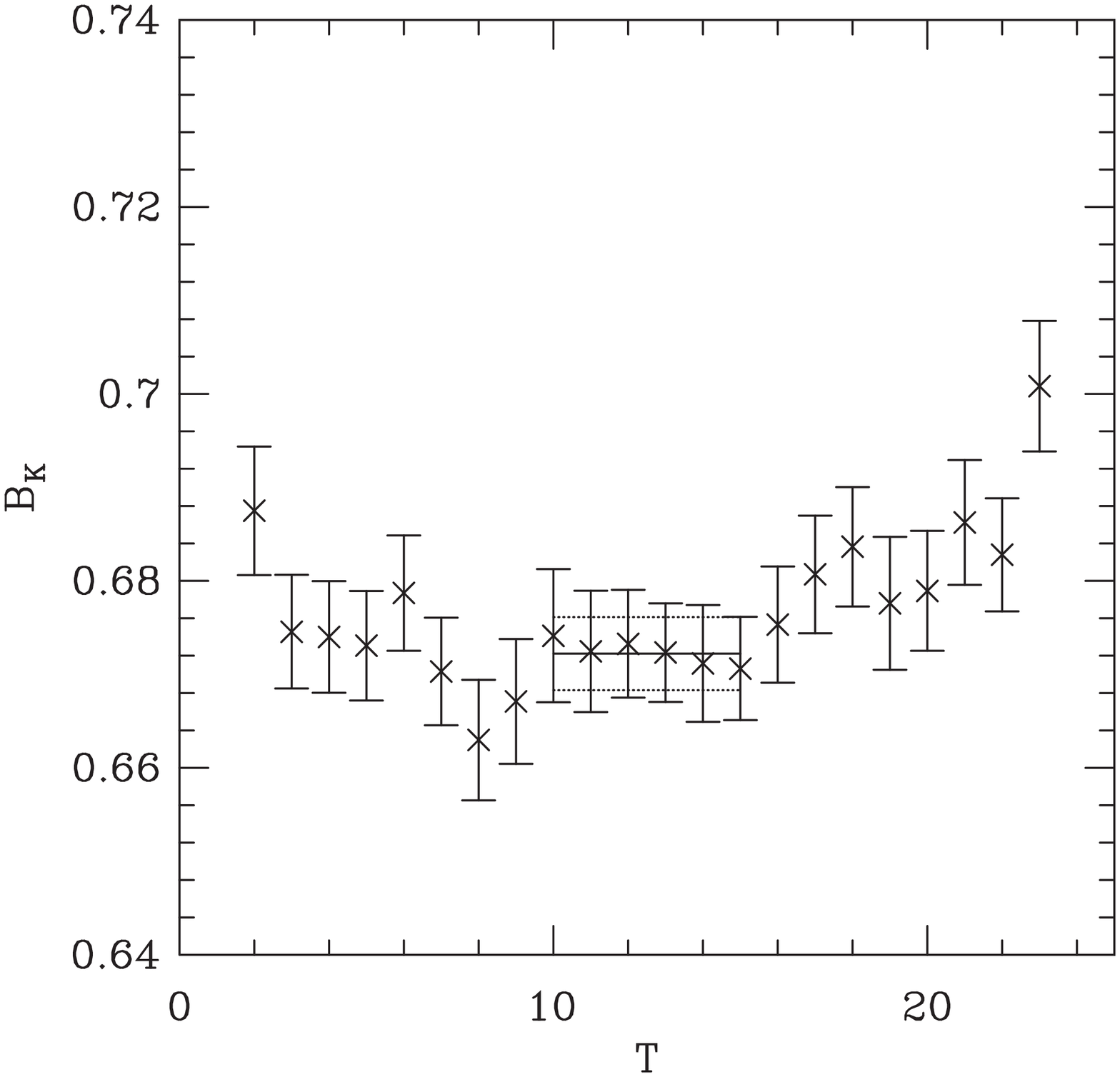, width=0.5\textwidth}
\epsfig{file=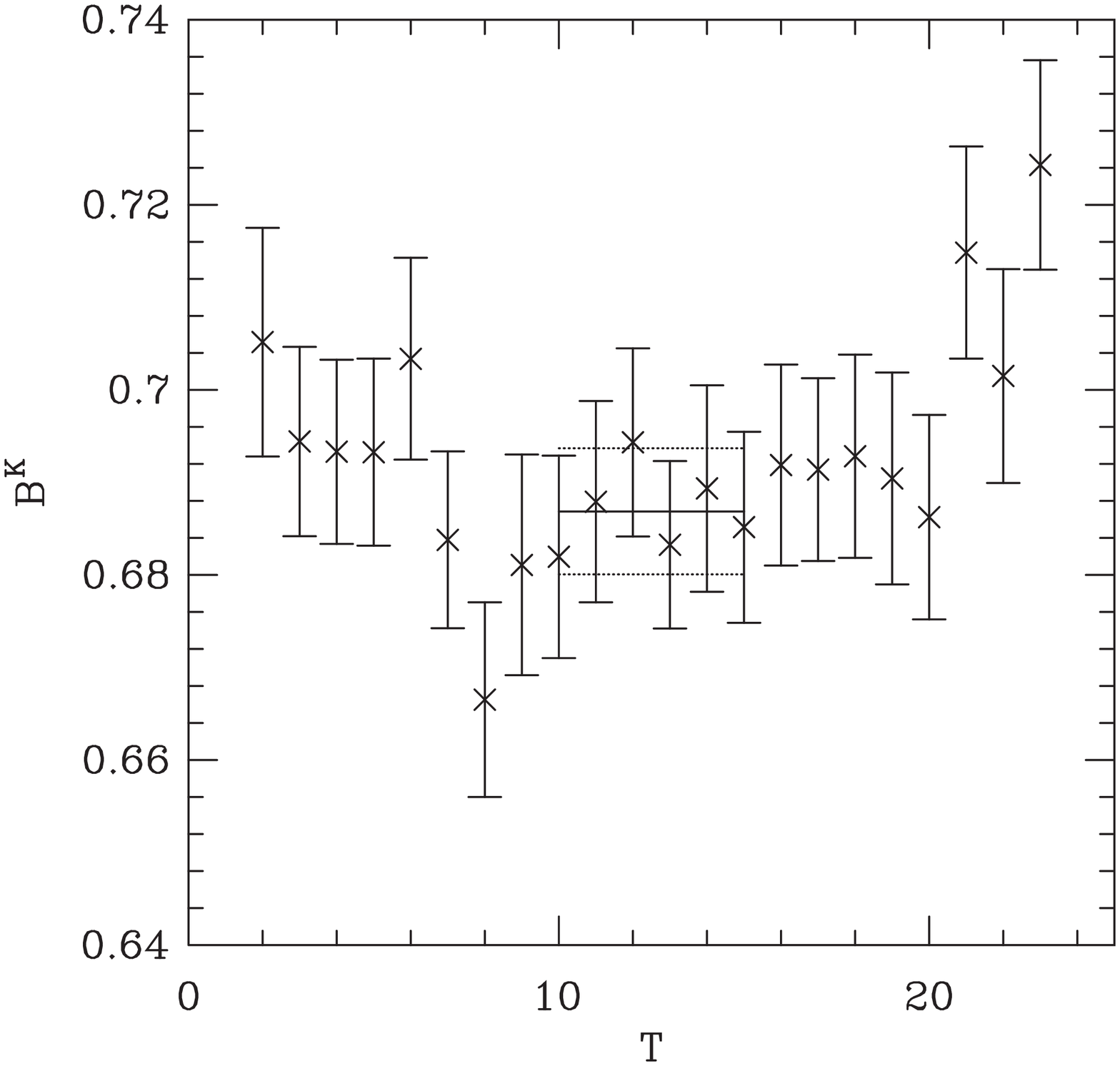, width=0.5\textwidth}
\caption{$B_K$ vs.~T for degenerate valence quarks with $m_x = m_y =
  0.03$ (left) and for non-degenerate valence quarks with $m_x = 0.01$
  and $m_y=0.05$ (right)}
\label{fig:bk-t}
\end{figure}

In Fig.~\ref{fig:bk-t}, we present $B_K$ as a function of Euclidean
time with a degenerate quark combination ($m_x= m_y=0.03$) in the
lefthand side and with a non-degenerate quark combination 
($m_x=0.01$ and $m_y=0.05$) in the righthand side.
The best fitting range is $10 \le t \le 15$ as explained in
Ref.~\cite{ref:wlee:10}.
Here, note that the quark masses are chosen such that the kaon masses
are the same at the leading order for both the degenerate and
non-degenerate combinations.
In Fig.~\ref{fig:bk-t}, we observe two things: (1) the signal for the
non-degenerate combination is noisier than that for the degenerate one
and (2) the $B_K$ value of the non-degenerate combination is
noticeably higher than that of the degenerate one.
In Ref.~\cite{ref:sharpe:1}, Van de Water and Sharpe calculated the
chiral behavior of $B_K$ using staggered chiral perturbation.
They provide results for the case of $N_f = 2+1$ partially quenched
QCD ($m_u = m_d \ne m_s$) in the continuum as well as at a finite
lattice spacing with staggered fermions.
For the degenerate combination ($m_x = m_y$), the chiral behavior of
$B_K$ is given in Ref.~\cite{ref:wlee:10} and we do not repeat it
here.
For the non-degenerate combination ($m_x \ne m_y$), they predict the
chiral behavior of $B_K$ in the continuum limit as follows:
\begin{eqnarray}
B_K &=& \tilde{c}_1 \bigg( 1 +
\frac{1}{48\pi^2 f^2} \Big[ I_{\rm conn} + I_{\rm disc} 
+ \tilde{c}_2 m_{xy}^2 
+ \tilde{c}_3 \frac{ (m_X^2-m_Y^2)^2 }{ m_{xy}^2 }
+ \tilde{c}_4 ( 2 m_U^2 + m_S^2 ) \Big]
\nonumber \\ & & \hspace*{+10mm}
+ \cdots \bigg)
\label{eq:bk:fit}
\end{eqnarray}
where $f=132$ MeV and $\tilde{c}_i$ are unknown dimensionless
low-energy constants.
The chiral logarithms are given by $I_{\rm conn}$ and
$I_{\rm disc}$.
The contribution from the quark connected diagrams is
\begin{eqnarray}
I_{\rm conn} &=& 6 m_{xy}^2 \tilde{l}(m_{xy}^2) 
- 3 l(m_X^2) \Big( 1 + \frac{m_X^2}{m_{xy}^2} \Big)
- 3 l(m_Y^2) \Big( 1 + \frac{m_Y^2}{m_{xy}^2} \Big)
\end{eqnarray}
The contribution from the diagrams involving a hairpin vertex is
\begin{eqnarray}
I_{\rm disc} &=& ( I_X + I_Y + I_\eta ) / m_{xy}^2
\\
I_X &=& \tilde{l}(m_X^2) 
\frac{(m_{xy}^2+m_X^2)(m_U^2-m_X^2)(m_S^2-m_X^2)}{m_\eta^2-m_X^2}
\nonumber \\ & &
- l(m_X^2) 
\frac{(m_{xy}^2+m_X^2)(m_U^2-m_X^2)(m_S^2-m_X^2)}{(m_\eta^2-m_X^2)^2}
\nonumber \\ & &
- l(m_X^2) 
\frac{2 (m_{xy}^2+m_X^2)(m_U^2-m_X^2)(m_S^2-m_X^2)}
{(m_Y^2-m_X^2)(m_\eta^2-m_X^2)}
\nonumber \\ & &
- l(m_X^2) 
\frac{ (m_U^2-m_X^2)(m_S^2-m_X^2)
-(m_{xy}^2+m_X^2)(m_S^2-m_X^2)
-(m_{xy}^2+m_X^2)(m_U^2-m_X^2)}
{ m_\eta^2-m_X^2 }
\\
I_Y &=& I_X( m_X^2 \leftrightarrow m_Y^2 )
\\
I_\eta &=& l(m_\eta^2) 
\frac{ (m_X^2-m_Y^2)^2(m_{xy}^2+m_\eta^2)(m_U^2-m_\eta^2)(m_S^2-m_\eta^2) }
{ (m_X^2-m_\eta^2)^2 (m_Y^2-m_\eta^2)^2 }
\end{eqnarray}
Here, note that the chiral logarithms appear through $l(X)$ and
$\tilde{l}(X)$, which are defined as
\begin{eqnarray}
l(X) &=& X \log( X/ \Lambda^2 ) + {\rm F.V.}
\\
\tilde{l}(X) &=& - [ \log( X / \Lambda^2 ) + 1 ] + {\rm F.V.} 
\end{eqnarray}
where ${\rm F.V.}$ represents a finite volume correction.
The notations for the various meson masses are as follows for those
composed of sea quarks:
\begin{eqnarray}
m_U^2 = 2 \mu m_d, \qquad 
m_S^2 = 2 \mu m_s, \qquad
m_\eta^2 = (m_U^2 + 2 m_S^2) / 3
\end{eqnarray}
and as follows for those composed of valence quarks
\begin{eqnarray}
m_X^2 = 2 \mu m_x, \qquad 
m_Y^2 = 2 \mu m_y, \qquad
m_{xy}^2 = \mu (m_x + m_y)
\end{eqnarray}
Here, we set sea quark masses to $m_u=m_d \ne m_s$ and
the two valence quark masses are $m_x$ and $m_y$. 
\begin{figure}[h!]
\begin{center}
\epsfig{file=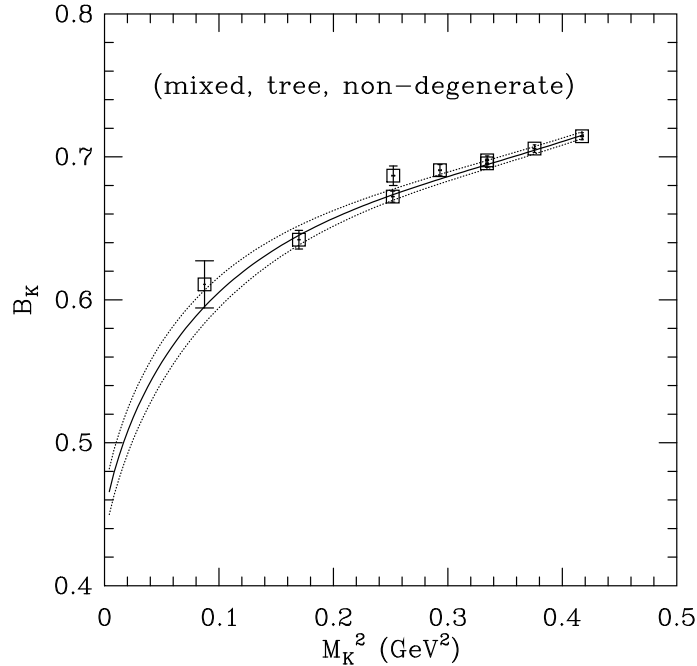, width=0.6\textwidth}
\end{center}
\caption{$B_K$ vs. $M_K^2$}
\label{fig:bk}
\end{figure}
In Fig.~\ref{fig:bk}, we plot $B_K$ data as a function of $M_K^2$.
Here, it includes both degenerate combinations ($m_x = m_y$) and
non-degenerate combinations ($m_x \ne m_y$).
We fit the data to the form of Eq.~(\ref{eq:bk:fit}) suggested by
chiral perturbation theory:
\begin{eqnarray}
B_K &=& c_1 \bigg( 1 +
\frac{1}{48\pi^2 f^2} \Big[ I_{\rm conn} + I_{\rm disc} \Big] \bigg)
+ c_2 m_{xy}^2
+ c_3 \frac{ (m_X^2 - m_Y^2)^2 }{ m_{xy}^2 }
+ c_4 m_{xy}^4
\end{eqnarray}
where the cut-off scale $\Lambda$ is set to $\Lambda=4 \pi f$.
The fitting results are summarized into Table \ref{tab:bk:fit}.
%
%
%
%
\begin{table}[h!]
\begin{center}
\begin{tabular}{| c | c | c |}
\hline
parameters & average & error \\
\hline
$c_1$           & 0.4469    & 0.0159 \\
$c_2$           & $-$1.4629 & 0.1788 \\
$c_3$           & 0.0216    & 0.0075 \\
$c_4$           & 1.4206    & 0.2133 \\
\hline
$\chi^2$/d.o.f. & 0.2999    & 0.2668 \\
\hline
\end{tabular}
\end{center}
\caption{Fitting results for $B_K$}
\label{tab:bk:fit}
\end{table}

When we compare the fitting results of Table \ref{tab:bk:fit} with
those in Ref.~\cite{ref:wlee:10} (only for the degenerate
combinations), we notice three things: (1) $\chi^2$ is much smaller
for the whole fitting (including both degenerate and non-degenerate
combinations) than that only for the degenerate combinations, (2) the
low-energy constants $c_1$, $c_2$ and $c_4$ are consistent with those
of Ref.~\cite{ref:wlee:10} within statistical uncertainty, and (3) for
a given choice of $\Lambda$, the $c_3$ term is negligibly small.
Judging from these observations, we understand that the effect of
non-degenerate quarks are dominated by the chiral logarithmic terms
such as $I_{\rm conn}$ and $I_{\rm disc}$ whereas the $c_3$ term
does not play much role at all.
Judging from the fact that the $c_3$ term is so small, we know that we
can determine all the low-energy constants ($c_1$, $c_2$, $c_4$) using
only degenerate combinations.
In addition, using these low-energy constants, we can determine the
$B_K$ value at the physical kaon mass with physical valence quarks
($m_x = m_d$ and $m_y = m_s$).
However, we know that $\chi^2$ is substantially smaller when we
include non-degenerate combinations in fitting, which leads us to the
conclusion that it would be significantly better for data analysis to
measure enough non-degenerate combinations as well as degenerate
combinations.

In Table \ref{tab:bk:fit}, the $\chi^2$ is rather high when we take
into account the fact that all the data are correlated.
In fact, one can observe that the fitting curve does not fit the
lightest two data points very well in Fig.~\ref{fig:bk}.
In other words, the fitting curve miss the lightest two data points in
the opposite direction.
In Ref.~\cite{ref:sharpe:1}, it is pointed out that the contribution
from the non-Goldstone pions are so significant that the curvature of
the fitting curve becomes smoother, which is consistent with what we
observe in Fig.~\ref{fig:bk}.
However, the full prediction from staggered chiral perturbation
contains 21 unknown low-energy constants for a single lattice spacing
and 37 unknown low-energy constants for the full analysis
\cite{ref:sharpe:1}.
In order to determine all of them, it is necessary to carry out a
significantly more extensive numerical work.

In the current analysis of $B_K$ data, we match the lattice results to
the continuum values at the tree level.
In this respect, the results are preliminary.

\section{Effect of non-degenerate quarks on $B_7^{(3/2)}$ and $B_8^{(3/2)}$}
\label{sec:b7:b8}
\begin{figure}[h!]
\epsfig{file=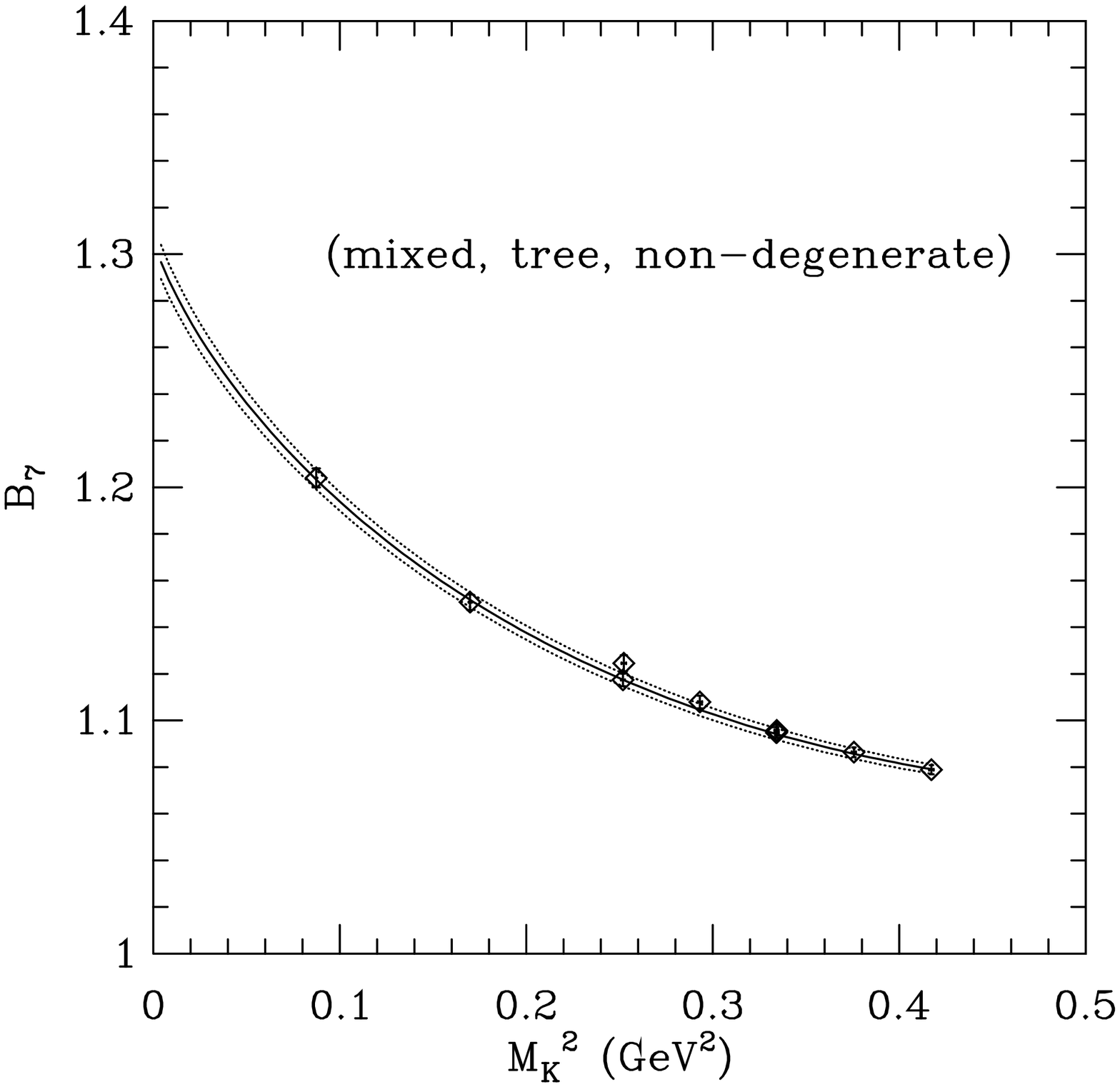, width=0.5\textwidth}
\epsfig{file=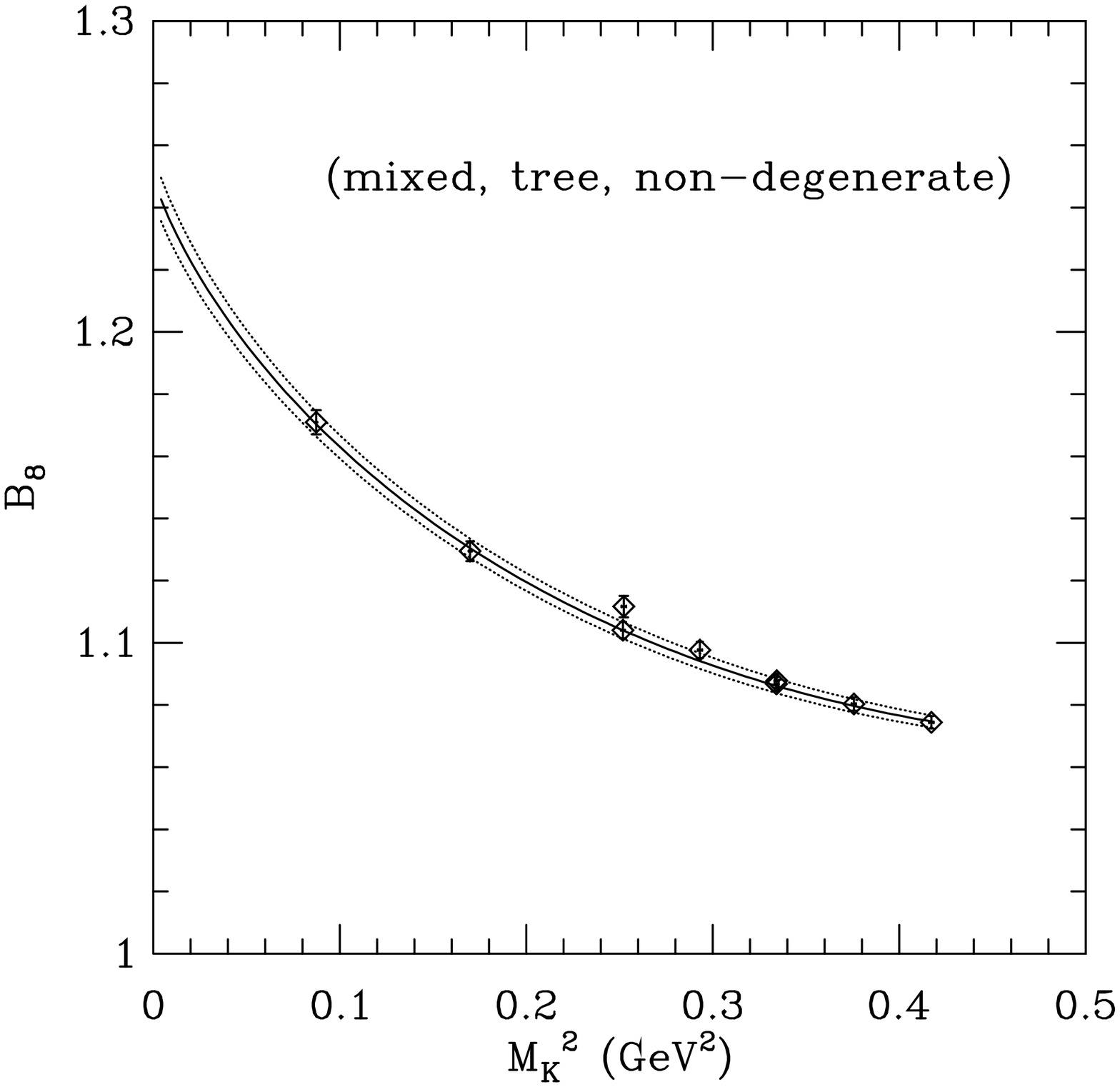, width=0.5\textwidth}
\caption{$B_7^{(3/2)}$ vs. $M_K^2$ (left) and $B_8^{(3/2)}$
vs. $M_K^2$ (right)}
\label{fig:b7:b8}
\end{figure}
In Fig.~\ref{fig:b7:b8}, we plot $B_7^{(3/2)}$ in the lefthand side
and $B_8^{(3/2)}$ in the righthand side as a function of $M_K^2$.
We show data of both degenerate and non-degenerate quark combinations.
The difference in $B_7^{(3/2)}$ ($B_8^{(3/2)}$) between the quark mass
pairs of (0.03, 0.03) and (0.01, 0.05) is about 0.6\% (0.7\%), which
is much smaller than that in $B_K$ (about 2\%).
Hence, the effect of non-degenerate quarks on $B_{7,8}^{(3/2)}$ is of
less interest as much.

There are some calculation on the chiral behavior of $\langle \pi^+ |
Q_{(8,8)}^{(3/2)} | K^+ \rangle $ available in
Ref.~\cite{ref:laiho:1}.
This result applies to the $N_f=3$ case with $m_u=m_d=m_s$.
It, however, does not apply to the case of $N_f = 2+1$ with $m_u = m_d
\ne m_s$.
Therefore, in order to understand the data in Fig.~\ref{fig:b7:b8}
further, it is necessary to calculate the chiral behavior for $N_f =
2+1$ using staggered chiral perturbation theory.
%


%
%
%


\begin{thebibliography}{99}
%
\bibitem{ref:wlee:10} Jongjeong Kim, Taegil Bae and Weonjong Lee,
\emph{Calculating $B_K$ using a mixed action}, in proceedings of
\emph{XXIIIrd International Symposium on Lattice Field Theory}, PoS
(LAT2005) 338, 
[{\tt hep-lat/0510007}].
%
\bibitem{ref:wlee:1} Weonjong Lee {\em et al.}, \emph{Testing improved
staggered fermions with $m_s$ and $B_K$}, 
\emph{Phys.~Rev.} D{\bf 71} (2005) 094501,
[{\tt hep-lat/0409047}].
%
\bibitem{ref:wlee:2} Weonjong Lee and Stephen Sharpe,
\emph{Perturbative matching of staggered four-fermion operators with
hypercubic fat links}, \emph{Phys.~Rev.}  D{\bf 68} (2003) 054510,
[{\tt hep-lat/0306016}].
%
\bibitem{ref:wlee:3} Weonjong Lee and Stephen Sharpe, \emph{One-loop
matching coefficients for improved staggered bilinears },
\emph{Phys.~Rev.}  D{\bf 66} (2002) 114501, [{\tt hep-lat/0208018}].
%
\bibitem{ref:milc:1} C. Aubin {\em et al.}, \emph{ Light hadrons with
improved staggered quarks: approaching the continuum limit },
\emph{Phys.~Rev.} D{\bf 70} (2004) 094505, [{\tt hep-lat/0402030}].
%
\bibitem{ref:sharpe:1} Ruth Van de Water and Stephen Sharpe, \emph{
$B_K$ in Staggered Chiral Perturbation Theory }, [{\tt
hep-lat/0507012}].
%
\bibitem{ref:laiho:1} J.~Laiho and A.~Soni, \emph{Lattice extraction of
$ K \to \pi \pi $ amplitudes to NLO in partially quenched
and in full chiral perturbation theory},
\emph{Phys.~Rev.} D{\bf 71} (2005) 014021,
[{\tt hep-lat/0306035}].
%
\end{thebibliography}
\end{document}